\documentclass[12pt]{iopart}
\newtheorem{theorem}{Theorem}
\begin{document}

\title[On the compatibility of Lorentz metrics with linear connections]
{On the compatibility of Lorentz metrics with linear connections on 4-dimensional manifolds}

\author{G S Hall\dag \ and D P Lonie\ddag\ }

\address{\dag Department of Mathematical Sciences, University of Aberdeen,
Meston Building, Aberdeen, AB24 3UE, Scotland, U.K. e-mail:
g.hall@maths.abdn.ac.uk }

\address{\ddag\ 108e Anderson Drive, Aberdeen, AB15 6BW, U.K. e-mail: DLonie@aol.com}

\begin{abstract}
This paper considers 4-dimensional manifolds upon which there is
 a Lorentz metric $h$ and a symmetric connection $\Gamma$ and
 which are originally assumed unrelated. It then derives
 sufficient conditions on $h$ and $\Gamma$ (expressed through the
 curvature tensor of $\Gamma$) for $\Gamma$ to be the Levi-Civita
 connection of some (local) Lorentz metric $g$ and calculates the
 relationship between $g$ and $h$. Some examples are provided
 which help to assess the strength of the sufficient conditions
 derived.
\end{abstract}
\submitto{\CQG} \pacs{04.20.-q, 04.20.Cv, 02.40.Ky } \maketitle

\section{Introduction}
If $M$ is a connected $n$-dimensional manifold admitting a smooth
metric $g$ of arbitrary signature, associated Levi-Civita
connection $\Gamma$ and corresponding curvature tensor $R$ then,
if $\nabla$ denotes the covariant derivative from $\Gamma$, one
has $\nabla g=0$ and, in components in any coordinate system,
\begin{equation}
g_{ae}R^e_{\ bcd}+g_{be}R^e_{\ acd}=0,\ \ \  g_{ae}R^e_{\
bcd;f}+g_{be}R^e_{\ acd;f}=0,\ \ ...
\end{equation}
where a semicolon denotes the covariant derivative $\nabla$ in
component form. This has led to the following question (or
variants of it) in the literature. Suppose a connected
$n$-dimensional manifold $M$ admits a smooth symmetric linear
connection $\Gamma$ with corresponding curvature tensor $R$ and
suppose, in addition, that $M$ admits a global smooth metric $h$
of arbitrary signature such that, on $M$
\begin{equation}
h_{ae}R^e_{\ bcd}+h_{be}R^e_{\ acd}=0
\end{equation}
\begin{flushright}
$ \begin{array}{lcc}
h_{ae}R^e_{\ bcd;f_1}+h_{be}R^e_{\ acd;f_1}=0 & \hspace{5.6cm} & (3.1) \hspace{-2mm}\\
\hspace{2cm}\vdots & \hspace{5.6cm} & \vdots\hspace{-2mm}\\
h_{ae}R^e_{\ bcd;f_1...f_k}+h_{be}R^e_{\
acd;f_1...f_k}=0 &\hspace{5.6cm} & (3.k)\hspace{-2mm}\\
\end{array}$
\end{flushright}
\addtocounter{equation}{1} for some integer $k$. Under which
conditions is $\Gamma$ a {\em metric (or locally metric)
connection}, that is, under which conditions does there exist a
global metric $g$ on $M$, of arbitrary signature, whose
Levi-Civita connection is $\Gamma$ (or, given $m\in M$, does
there exist an open neighbourhood $U$ of $m$ and metric $g$ of
arbitrary signature on $U$ whose Levi-Civita connection is the
restriction to $U$ of $\Gamma$) and, if so, how are $g$ and $h$
related?

If one is somehow able to find the holonomy group of $\Gamma$
then the problem is partly solved because then $\Gamma$ is a
metric connection if and only if for some $m\in M$ there is a
non-degenerate quadratic form  on the tangent space $T_mM$ to $M$
at $m$, of signature $(p,q)$ for non-negative integers $p, q$ with
$p+q=n$, which is preserved by the holonomy group associated with
$\Gamma$. If such is the case, $\Gamma$ is compatible with a
metric $g$ on $M$ of signature $(p, q)$, that is $\nabla g=0$, and
the holonomy group of $\Gamma$ is a Lie-subgroup of $O(p, q)$
\cite{K&N,Schmidt}. In the case $n=4$ and if $\Gamma$ is fixed
and given to be the Levi-Civita connection of a Lorentz metric on
$M$, the holonomy group of $\Gamma$ can be used to find {\em all}
metrics on $M$ compatible with $\Gamma$ and these may (depending
on the holonomy group) have any of the signatures $(1,3)$,
$(0,4)$ or $(2,2)$ \cite{Hall1988}. Remaining in the $n=4$ case
and with $R$ fixed and assumed to arise from a Levi-Civita
connection $\Gamma$ compatible with a Lorentz metric on $M$, it
is known that this metric is generically determined up to a
constant conformal factor and hence, generically, $\Gamma$ is
uniquely determined \cite{HallBook,Hall1983,Ihrig,Rendall1988}. It
is also known that on a large class of 4-dimensional manifolds a
symmetric linear connection $\Gamma$ satisfying (2) for some
metric $h$ on $M$ is necessarily {\em locally} metric (in the
sense that each $m\in M$ admits an open neighbourhood $U$ such
that the restriction of $\Gamma$ to $U$ is metric). However, the
procedure involves the investigation of a $36\times 36$ matrix
and is geometrically obscure \cite{Edgar}. In this paper, the
question raised at the beginning of this section for the case
$n=4$ and with $h$ of Lorentz signature will be considered, that
is, if $M$ admits a symmetric linear connection $\Gamma$,
associated curvature $R$ and metric tensor $h$ of Lorentz
signature (and {\em not} assumed related to $\Gamma$ or $R$ in
any way) such that (2) and (3.1)-(3.$k$) hold for some $k$, is
$\Gamma$ a metric (or locally metric) connection on $M$ and, if
so, how is $h$ related to the metric or metrics compatible with
$\Gamma$?

A standard notation will be used, with round and square brackets
denoting the usual symmetrisation and and skew-symmetrisation of
indices, respectively, and a comma denotes a partial derivative.
A skew-symmetric tensor $F$ of type $(0,2)$ or $(2,0)$ at $m$ is
called a {\em bivector}. If $F(\neq 0)$ is such a bivector, the
rank of any of its (component) matrices is either two or four. In
the former case, one may write (e.g. in the $(2,0)$ case)
$F^{ab}=2r^{[a}s^{b]}$ for $r, s\in T_mM$ (or alternatively,
$F=r\wedge s$) and $F$ is called {\em simple}, with the
2-dimensional subspace (2-space) of $T_mM$ spanned by $r, s$
referred to as the {\em blade} of $F$. In the latter case, $F$ is
called {\em non-simple}.

The metric $h(m)$ converts $T_mM$ into a Lorentz inner product
space and thus it makes sense to refer to vectors in $T_mM$ and
covectors in the cotangent space $T_m^{*}M$ to $M$ at $m$ (using
$h(m)$ to give a unique isomorphism $T_mM\leftrightarrow
T_m^{*}M$, that is, to raise and lower tensor indices) as being
{\em timelike}, {\em spacelike}, {\em null} or {\em orthogonal},
using the signature $(-,+,+,+)$. The same applies to 1-dimensional
subspaces ({\em directions}) and 2- and 3-dimensional subspaces of
$T_mM$ or $T^{*}_mM$. A {\em simple} bivector at $m$ is then
called {\em timelike} (respectively, {\em spacelike} or {\em
null}) if its blade at $m$ is a timelike (respectively a
spacelike or null) 2-space at $m$. A {\em non-simple} bivector
$F$ at $m$ may, by a standard argument (\cite{Sachs}, see also
\cite{HallBook}), be shown to uniquely determine an orthogonal
pair of 2-spaces at $m$, one spacelike and one timelike, and
which are referred to as the {\em canonical pair of blades} of
$F$. A tetrad $(l,n,x,y)$ of members of $T_mM$ is called a null
tetrad at $m$ if the only non-vanishing inner products between
its members at $m$ are $h(l,n)=h(x,x)=h(y,y)=1$. Thus $l$ and $n$
are null.

It is also remarked that the tensor, with local components
$R_{abcd}\equiv h_{ae}R^e_{\ bcd}$ satisfies
$R_{abcd}=-R_{bacd}=-R_{abdc}$ and
$R_{abcd}+R_{adbc}+R_{acdb}=0$. It then follows after some index
juggling (see, e.g. \cite{Adler}), that $R_{abcd}=R_{cdab}$.

\section{Preliminary Results}
Let $M$ be a 4-dimensional smooth connected Hausdorff manifold
admitting a smooth symmetric linear connection $\Gamma$ with
associated curvature tensor $R$ and a global smooth Lorentz
metric $h$ such that (2) holds. No relation between $h$ and
$\Gamma$ is assumed other than (2). It will be convenient to
describe a simple algebraic classification of $R$ (relative to
$h$). This classification is easily described geometrically and is
a pointwise classification. It has been described before in a
more specific context (see, e.g. \cite{HallBook,Hall&Lonie2004})
but since its use here is slightly different, it will be briefly
described.

Define a linear map $f$ from the 6-dimensional vector space of
type $(2,0)$ bivectors at $m$ into the vector space of type
$(1,1)$ tensors at $m$ by $f:F^{ab}\rightarrow R^a_{\
bcd}F^{cd}$. The condition (2) shows that if a tensor $T$ is in
the range of $f$ then
\begin{equation}
h_{ae}T^e_{\ b}+h_{be}T^e_{\ a}=0 \ \ \ \ (\Rightarrow
T_{ab}=-T_{ba}, \ \  T_{ab}=h_{ae}T^e_{\ b})
\end{equation}
and so $T$ can be regarded as a member of the matrix
representation of the Lie algebra of the pseudo-orthogonal
(Lorentz) group of $h(m)$. Using $f$ one can divide the curvature
tensor $R(m)$ into five classes.

\begin{description}
\item[Class $A$] This is the most general curvature class and the curvature
will be said to be of (curvature) class $A$ at $m\in M$ if it is
not in any of the classes $B$, $C$, $D$ or $O$ below.
\item[Class $B$] The curvature tensor is said to be of (curvature)
class $B$ at $m\in M$ if the range of
$f$ is 2-dimensional and consists of all linear combinations of
type $(1,1)$ tensors $F$ and $G$ where $F^a_{\ b}=x^ay_b-y^ax_b$
and $G^a_{\ b}=l^an_b-n^al_b$ with $l,n,x,y$ a null tetrad at
$m$. The curvature tensor at $m$ can then be written as
\begin{equation}
  R_{abcd}\equiv h_{ae}R^e_{\ bcd}=\frac{\alpha}{2} F_{ab}F_{cd}-\frac{\beta}{2} G_{ab}G_{cd}
\end{equation}
where $\alpha,\beta\in\mathbf{R}$ $\alpha\neq 0\neq\beta$, the
symmetrised cross term in $F$ and $G$ vanishes because
$R_{a[bcd]}=0$. This class can be further split in classes $B_1$
and $B_2$ where $B_1$ (respectively, $B_2$) applies if
$\alpha\neq\beta$ (respectively, $\alpha=\beta$).
\item [Class $C$] The
curvature tensor is said to be of (curvature) class $C$ at $m\in
M$ if the range of $f$ is 2- or 3-dimensional and if there exists
$0\neq k\in T_mM$ such that each of the type $(1,1)$ tensors in
the range of $f$ contains $k$ in its kernel (i.e. each of their
matrix representations $F$ satisfies $F^a_{\ b}k^b=0$).
\item[Class $D$] The curvature tensor is said to be of (curvature) class $D$ at $m\in M$ if the range of
$f$ is 1-dimensional. It follows that the curvature components
satisfy $R_{abcd}=\lambda F_{ab}F_{cd}$ at $m$ $(0\neq\lambda\in
\mathbf{R})$ for some bivector $F$ at $m$ which then satisfies
$F_{a[b}F_{cd]}$ and is thus simple. It also follows that there
exist two independent members $r,s\in T_mM$ such that
$F_{ab}r^b=F_{ab}s^b=0$ and hence that $r$ and $s$ lie in the
kernel of each tensor in the range of $f$.
\item[Class $O$] The curvature tensor is said to be of (curvature) class $O$ at $m\in M$ if it vanishes at
$m$.
\end{description}
The following remarks can be checked in a straightforward manner
\cite{HallBook}.
\begin{enumerate}
\item For the classes $A$ and $B$ there does {\bf not} exist
$0\neq k\in T_mM$ such that $F^a_{\ b}k^b=0$ for {\em every} $F$
in the range of $f$.
\item For class $A$, the range of $f$ has
dimension at least two and if this dimension is four or more the
class is necessarily $A$.
\item The vector $k$ in the definition of class $C$ is
unique up to a scaling.
\item For the classes $A$ and $B$ there
does {\bf not} exist $0\neq k\in T_mM$ such that $R^a_{\
bcd}k^d=0$, whereas this equation has exactly one independent
solution for class $C$ and two for class $D$.
\item The five classes $A$, $B$, $C$, $D$ and $O$ are mutually
exclusive and exhaustive for the curvature tensor at $m$. If the
curvature class is the same at each $m\in M$ then $M$ will be
said to be of that class.
\end{enumerate}

Denoting by the same symbols $A$, $B$ (or $B_1$ and $B_2$), $C$ ,
$D$ and $O$ those subsets of $M$ consisting of precisely those
points where the curvature tensor has that class, one may
decompose $M$, disjointly as $M=A\cup B_1\cup B_2\cup C\cup D\cup
O$ and then, disjointly, as
\begin{equation}
M=A\cup\mathrm{int}B_1\cup\mathrm{int}B_2\cup\mathrm{int}C\cup\mathrm{int}D\cup\mathrm{int}O\cup
Z
\end{equation}
where $\mathrm{int}$ denotes the manifold topology interior
operator and $Z$ is the (necessarily closed) subset of $M$
defined by the disjointness. The idea is to show that $Z$ has
empty interior, $\mathrm{int}Z=\emptyset$. A similar
decomposition was considered in
\cite{HallBook,Hall&Lonie2004,Patel}, but in these works $B$ was
not subdivided into $B_1$ and $B_2$ and so the "leftovers" subset
$Z$ may differ from that in (6) since
$\mathrm{int}B_1\cup\mathrm{int}B_2$ may be a proper subset of
$\mathrm{int}B$. To show that $\mathrm{int}Z=\emptyset$ in (6) it
is first noted that $A$ is open in $M$ (i.e. $\mathrm{int}A=A$)
and that $A\cup B$ is also open in $M$ \cite{HallBook}. So let
$U\subseteq Z$ be open in $M$. Then by the disjointness of the
decomposition (6) $U\cap A=\emptyset$ and so $U\cap(A\cup
B)=U\cap B\equiv W$ and $W$ is open in $M$. The disjointness of
(6) also shows that, if $W\neq\emptyset$, $W$ must intersect each
of $B_1$ and $B_2$ non-trivially, so let $m\in W\cap B_1$ and
consider the linear map $f'$ from the vector space of type
$(2,0)$ bivectors at $m$ into itself given by
$f':F^{ab}\rightarrow R^{ab}_{\ \ cd}F^{cd}$. The characteristic
polynomial of this map has, from (5), three distinct solutions
$\alpha$, $\beta$ and $0$ at $m$ and since the first two of these
are {\em simple} roots they give rise to smooth functions
$\tilde{\alpha}$ and $\tilde{\beta}$ defined on some
neighbourhood $V$ of $m$ (and with $\tilde{\alpha}(m)=\alpha$ and
$\tilde{\beta}(m)=\beta$) which are solutions of the
characteristic polynomial of $f'$ and distinct from each other
and from zero in $V$ \cite{Hall&Rendall1989}. It follows that
$m\in V\cap W\subseteq B_1$ with $V\cap W$ open and disjoint from
$B_2$ (since at points of $B_2$ the characteristic polynomial of
$f'$ has only two distinct solutions $\alpha (=\beta)$ and $0$).
Thus $V\cap W\subseteq\mathrm{int}B_1$ and hence
$U\cap\mathrm{int}B_1\neq\emptyset$, contradicting the
disjointness of (6). It follows that $W$ and hence $U\cap B$ are
empty. From here the argument follows \cite{HallBook} to get
$U=\emptyset$ and hence $\mathrm{int}Z=\emptyset$. The following
is thus established.
\begin{theorem}
In the disjoint decomposition (6), $Z$ is closed,
$\mathrm{int}Z=\emptyset$ and so $M\backslash Z$ is an open dense
subset of $M$.
\end{theorem}
It is remarked that the reason for decomposing $B$ as $B_1\cup
B_2$ is that the above argument leading to the local smooth
functions $\tilde{\alpha}$ and $\tilde{\beta}$ and which extend
$\alpha$ and $\beta$ does not apply if $\alpha=\beta$ at $m$
unless $m\in \mathrm{int}B_2$, (i.e. the problem lies with the
points in $B_2\setminus\mathrm{int}B_2$).

It is also remarked that in the open subset $\mathrm{int}C$ of $M$
the non-zero tangent vectors $k$ introduced in the definition of
this type give rise to a {\em smooth} integrable 1-dimensional
distribution on $\mathrm{int}C$. To see this one recalls that the
curvature tensor satisfies $R^a_{\ bcd}k^d=0$ at each $m\in
\mathrm{int}C$ and that $k$ is unique (up to a scaling) in
satisfying this equation. The smoothness of the distribution
defined follows from the smoothness of the curvature in the
following way \cite{Hall&daCosta1988,HallBook}. Let $m\in
\mathrm{int}C$ and let $0\neq k\in T_mM$ satisfy the above
equation. Choose a coordinate domain $U\subseteq\mathrm{int}C$
about $m$ in which $k^4=1$ at $m$. The conditions $R^a_{\
bcd}k^d=0$ on $U$ then reduce to a system of equations of the form
$\sum^3_{\alpha=1}E_{p\alpha}k^{\alpha}=E_p$ for functions
$E_{p\alpha}$ and $E_p:U\rightarrow\mathbf{R}$ and $p=1,2,3$.
Since the linear system so formed has rank equal to three at $m$
(since there the $k^{\alpha}$ are uniquely determined) then by
taking $k^4=1$ on $U$ one obtains a linear system for the
$k^{\alpha}$ of rank three over $U$. Cramer's rule then reveals
smooth solutions for the components of $k^{\alpha}$ and hence a
local nowhere zero vector field $k$ satisfying $R^a_{\ bcd}k^d=0$.

In the open subset $\mathrm{int}D$ it can be shown that for
$m\in\mathrm{int}D$ there exists an open coordinate neighbourhood
$U\subseteq\mathrm{int}D$ of $m$ on which the curvature tensor
satisfies $R_{abcd}=\lambda F_{ab}F_{cd}$ where $F$ is a smooth
bivector and $\lambda$ a smooth real-valued function on $U$. A
similar argument to the previous one shows that $U$ may be chosen
so that there are smooth vector fields $r$ and $s$ on $U$ such
that at each $m'\in U$, $r(m')$ and $s(m')$ are independent and
satisfy $R^a_{\ bcd}r^d=R^a_{\ bcd}s^d=0$ \cite{Hall&Rendall1989}.

If $m\in\mathrm{int}B_1$ the curvature tensor takes the form (5)
over some coordinate neighbourhood $U\subseteq\mathrm{int}B_1$ of
$m$ with $\alpha,\beta$ smooth nowhere zero, nowhere equal
real-valued functions and $F$ and $G$ smooth bivectors on $U$.
Similar remarks apply if $m\in\mathrm{int}B_2$ with
$\alpha=\beta$ on $U\subseteq\mathrm{int}B_2$. For the latter two
($B_1$ and $B_2$) cases, a null tetrad $l,n,x,y$ of smooth vector
fields exists on $U$ satisfying (5) on $U$.

\section{The Main Results}
Let $M$, $\Gamma$, $R$ and $h$ be as described at the beginning
of section 2 and suppose that (2) holds. If (3.1) to (3.$k$) also
hold then it follows that on $M$
\begin{flushright}
$ \begin{array}{lcc}
h_{ae;f_1}R^e_{\ bcd}+h_{be;f_1}R^e_{\ acd}=0 & \hspace{5.6cm} & (7.1) \hspace{-2mm}\\
\hspace{2cm}\vdots & \hspace{5.6cm} & \vdots\hspace{-2mm}\\
h_{ae;f_1...f_k}R^e_{\ bcd}+h_{be;f_1...f_k}R^e_{\
acd}=0 &\hspace{5.6cm} & (7.k)\hspace{-2mm}\\
\end{array}$
\end{flushright}
\addtocounter{equation}{1} and conversely the set of equations
(2) and (7.1) to (7.$k$) imply the set (2) and (3.1) to (3.$k$).
Thus the conditions (2) and (3.1) to (3.$k$) are equivalent to the
conditions (2) and (7.1) to (7.$k$). The aim is to use this
equivalence to show when the original conditions (2) and (3.1) to
(3.$k$) for a particular $k$ imply the existence of a local or
global Lorentz metric $g$ on $M$ compatible with $\Gamma$ and to
display the relationship between $g$, $h$ and the geometry of $R$
as expressed through its curvature type. The method to be used
employs a theorem in \cite{Hall&McIntosh1983,HallBook} and
involves the following idea. Let $m\in M$ and $T$ be a type
$(0,a)$ tensor at $m$ with $a\geq 2$ and with $T$ symmetric in
its first two indices (so $T$ has components $T_{abc...d}$ and
$T_{abc...d}=T_{bac...d}$). Suppose also that at $m$ (cf. (2))
\begin{equation}
T_{aeg...h}R^e_{\ bcd}+T_{beg...h}R^e_{\ acd}=0
\end{equation}
Then for any tensor $S$ of type $(a-2,0)$, the tensor
$h'_{ab}=T_{abc...d}S^{c...d}$ is symmetric and satisfies (2).
This means that for any $F$ in the range of the map $f$ at $m$
described earlier
\begin{equation}
h'_{ae}F^e_{\ b}+h'_{be}F^e_{\ a}=0
\end{equation}
Each such $F$ in the range of $f$ imposes strong algebraic
constraints on $h'$, these constraints being conveniently written
down for each of the curvature classes for $R$ at $m$ in terms of
the original Lorentz metric $h$ and the features of the particular
curvature class. They are, in the notation of section 2
\cite{Hall&McIntosh1983,HallBook}
\begin{flushright}
$ \begin{array}{llcr}
\mathrm{Class\ A}\hspace{1cm} & h'_{ab}=\alpha h_{ab} & \hspace{1cm} & (10a)\hspace{-2mm} \\
\mathrm{Class\ B}\hspace{1cm} & h'_{ab}=\alpha h_{ab}+2\beta
l_{(a}n_{b)}=(\alpha+\beta)h_{ab}-\beta(x_ax_b+y_ay_b) & \hspace{1cm}& (10b)\hspace{-2mm} \\
\mathrm{Class\ C}\hspace{1cm} & h'_{ab}=\alpha h_{ab}+\beta k_ak_b
& \hspace{1cm}&
(10c)\hspace{-2mm} \\
\mathrm{Class\ D}\hspace{1cm} & h'_{ab}=\alpha h_{ab}+\beta
r_ar_b+\gamma
s_as_b+2\delta r_{(a}s_{b)} & \hspace{1cm}& (10d)\hspace{-2mm} \\
\end{array}$
\end{flushright} \addtocounter{equation}{1} where $\alpha, \beta,
\gamma, \delta\in\mathbf{R}$ and the completeness relation
$h_{ab}=2l_{(a}n_{b)}+x_ax_b+y_ay_b$ was used in (10b). These
expressions allow, depending on the curvature class, expressions
for the covariant derivatives of $h$ to be written out reasonably
conveniently since, from (2) and (7.1) to (7.$k$), they satisfy
the conditions asked for in (8). Hence, for any tensor $S$ of the
appropriate type, each tensor $h'_{ab}=h_{ab;c...d}S^{c...d}$
(for the appropriate number of derivatives) satisfies (9) for each
such $F$ and hence the appropriate equation in the set (10a) to
(10d). From the arbitrariness of $S$ one obtains the desired
expressions for these covariant derivatives. For example, if the
curvature class at $m$ is $C$ then, at $m$, and if (7.1) and
(7.2) are assumed
\begin{equation}
h_{ab;c}=h_{ab}\alpha_c+k_ak_b\beta_c, \ \ \ \ \
h_{ab;cd}=h_{ab}\alpha_{cd}+k_ak_b\beta_{cd}
\end{equation}
for covectors $\alpha_c$ and $\beta_c$ and tensors $\alpha_{cd}$
and $\beta_{cd}$ at $m$.

These ideas can now be applied to the open subsets $A$,
$\mathrm{int}B_1$, $\mathrm{int}B_2$, $\mathrm{int}C$ and
$\mathrm{int}D$ in the decomposition (6).
\begin{theorem}
Let $M$ be a smooth 4-dimensional connected manifold admitting a
smooth symmetric linear connection $\Gamma$ and associated
curvature $R$ and also admitting a smooth Lorentz metric $h$ such
that (2) holds. Then
\begin{enumerate}
  \item if, on $A$, (3.1) also holds, $\Gamma$ is compatible with
  a local Lorentz metric on $A$,
  \item if, on $\mathrm{int}B_1$ (respectively $\mathrm{int}B_2$), (3.1) also holds, $\Gamma$ is compatible with
  a local Lorentz metric on $\mathrm{int}B_1$ (respectively
  $\mathrm{int}B_2$),
  \item if, on $\mathrm{int}C$, (3.1) and (3.2) also hold, $\Gamma$ is compatible with
  a local Lorentz metric on an open dense subset of $\mathrm{int}C$,
   \item if, on $\mathrm{int}D$, (3.1), (3.2) and (3.3) also hold, $\Gamma$ is compatible with
  a local Lorentz metric on an open dense subset of $\mathrm{int}D$.
\end{enumerate}
\end{theorem}
{\bf Proof}\\
{\em (i)} The remarks preceding the theorem (see (10a)) together
with the imposition of (3.1) or, equivalently, (7.1) show that
$h_{ab;c}=h_{ab}w_c$ for some 1-form $w$ on the open subset $A$
which is easily seen to be smooth since $h$ and $\Gamma$ are. Now
use (2) and the Ricci identity on $h$ to get
\begin{equation}
h_{ab;[cd]}=0 \hspace{1cm} (\Rightarrow h_{ab}w_{[c;d]}=0)
\end{equation}
Thus $w_{[c;d]}=0$ and so $w_a$ is locally a gradient. Hence, for
each $m\in A$, there is an open neighbourhood $W(\subseteq A)$ of
$m$ on which $w_a=w_{,c}$ for some smooth function
$w:W\rightarrow\mathbf{R}$. Then on $W$, $g_{ab}=e^{-w}h_{ab}$
satisfies $g_{ab;c}=0$ and is a local Lorentz metric for
$\Gamma$. Further, if $g'$ is any other local metric defined on
some neighbourhood $W'$ of $m$ and compatible with $\Gamma$ then
$g'$ satisfies (2) on $W'$ and hence, on $W\cap W'$, $g'=\phi g$
for some positive smooth function $\phi:W\cap
W'\rightarrow\mathbf{R}$ (see (10a)). From this and the result
$g'_{ab;c}=0$ it follows that $g'$ is a constant multiple of $g$
on $W\cap
W'$.\\
\\
{\em (ii)} The remarks preceding the theorem (see (10b)) together
with the imposition of (3.1) and (7.1) show that for each
$m\in\mathrm{int}B_1$ (respectively, $m\in\mathrm{int}B_2$) there
is a coordinate neighbourhood $U\subseteq \mathrm{int}B_1$
(respectively, $U\subseteq \mathrm{int}B_2$) such that on $U$
\begin{equation}
h_{ab;c}=h_{ab}w_c+2l_{(a}n_{b)}\lambda_c
\end{equation}
where $l$ and $n$ are the smooth null members of the null tetrad
$l,n,x,y$ on $U$ established in section 2 and $w$ and $\lambda$
are (necessarily) smooth 1-forms on $U$. Since the contravariant
components of $h$ satisfy $h^{ac}h_{cb}=\delta^a_{\ b}$, the
result $(h^{ac}h_{cb})_{;c}=0$ implies that
\begin{equation}
h^{ab}_{\ \ ;c}=-h^{ab}w_c-2l^{(a}n^{b)}\lambda_c
\end{equation}
Now $U$ can be chosen so that, in addition to the above results,
the curvature tensor satisfies (5) on $U$ with $\alpha, \beta$
smooth nowhere zero functions and $F$ and $G$ as given in (5) in
terms of the null tetrad on $U$. The Bianchi identities between
$R$ and $\Gamma$ are $R^a_{\ b[cd;e]}=0$. One now employs a
technique first used in \cite{Hall&Kay1988} (where $h$ {\em was},
in fact, a metric compatible with $\Gamma$). First, one writes out
the Bianchi identity using (5) and contracts with
$l_ax^bl^cn^dx^e$. This operation eliminates all terms except one
and gives, on $U$
\begin{equation}
\beta G^a_{\ b;e}G_{cd}l_ax^bl^cn^dx^e=0 \hspace{1cm} (\Rightarrow
l_{a;b}x^ax^b=0)
\end{equation}
Similar contractions with $l_ax^bl^cn^dy^e$, $l_ax^bx^cy^dl^e$
and $l_ax^bx^cy^dn^e$ (and the result $l^ay_{a;b}=-y^al_{a;b}$
which follows from applying (14) to the expansion
    of $(h^{ab}l_ay_b)_{;c}=0$) lead to
\begin{equation}
l_{a;b}x^ay^b=l_{a;b}y^al^b=l_{a;b}y^an^b=0
\end{equation}
Observations of the symmetry in these contractions then give
\begin{equation}
l_{a;b}y^ay^b=l_{a;b}y^ax^b=l_{a;b}x^al^b=l_{a;b}x^an^b=0
\end{equation}
From these results it easily follows that
$l_{a;b}x^a=l_{a;b}y^a=0$. Further, one has
$(h_{ab}l^al^b)_{;c}=0$ and so, using (13), one has $l^a_{\
;b}l_a=0$ and hence $l_{a;b}l^a=0$. It follows that
$l_{a;b}=l_ap_b$ for some smooth 1-form $p$ on $U$ and then, from
(14), that $l^a_{\ ;b}=l^a(p_b-w_b-\lambda_b)$. Thus $l^a$ and
$l_a$ are {\em recurrent} on $U$. Similar arguments show that
$n^a$ and $n_a$ are {\em recurrent} on $U$ and, since
$(l^an_a)_{;b}=0$, that $n_{a;b}=n_a(\lambda_b+w_b-p_b)$ and
$n^a_{\ ;b}=-n^ap_b$. It is then easily checked that the nowhere
zero symmetric tensor $T_{ab}=2l_{(a}n_{b)}$ on $U$ is also
recurrent in the sense that $T_{ab;c}=T_{ab}r_c$ where $r$ is the
1-form $r_a=\lambda_a+w_a$ on $W$. When the Ricci identity is
applied to $T_{ab}$ and use is made of (5) one gets
\begin{equation}
T_{ab;[cd]}=2n_el_{(a}R^e_{\ b)cd}+2l_en_{(a}R^e_{\ b)cd}=0
\end{equation}
and so $r_{[a;b]}=0$. Thus one may assume, by shrinking $W$ if
necessary, that, on $W$, $r_a=r_{,c}$ for a smooth function
$r:W\rightarrow\mathbf{R}$. It follows that, on $W$, the tensor
$t_{ab}=e^{-r}T_{ab}$ is symmetric, nowhere zero and covariantly
constant, $t_{ab;c}=0$. When (13) is rewritten as
$h_{ab;c}=h_{ab}w_c+t_{ab}\gamma_c$ for some smooth 1-form
$\gamma_a$ on $W$ and the Ricci identity is applied to $h$ (using
(2)) one finds
\begin{equation}
h_{ab}w_{[c;d]}+t_{ab}(\gamma_{[c;d]}-\gamma_{[c}w_{d]})=0
\end{equation}
At each point of $W$ the matrices $h_{ab}$ and $t_{ab}$ have
(different) ranks four and two, respectively, and so the two
quantities on the left hand side of (19) must vanish separately.
The first of these equations leads, as before, to the fact that
$w_a$ is locally a gradient and the second gives
$\gamma_{[c;d]}-\gamma_{[c}w_{d]}=0$. Thus by again shrinking $W$
if necessary one has $w_a=w_{,a}$ on $W$ and then the second
condition can be rewritten as
\begin{equation}
(e^{-w}\gamma_a)_{;b}-(e^{-w}\gamma_b)_{;a}=0
\end{equation}
This means, by again reducing $W$ if necessary, that there exists
a smooth function $\delta:W\rightarrow\mathbf{R}$ such that
$\gamma_a=e^w\delta_{,a}$. Now consider a tensor $g$ on $W$ of the
form
\begin{equation}
  g_{ab}=\phi h_{ab}+\epsilon t_{ab}
\end{equation}
for smooth functions $\phi,\epsilon:W\rightarrow\mathbf{R}$ with
$\phi$ nowhere zero. Then, by again considering rank on the
indices $a,b$ one sees that the condition that $g$ is covariantly
constant, $g_{ab;c}=0$, is equivalent to the two differential
equations
\begin{equation}
  \phi_{,a}+\phi w_{,a}=0\hspace{2cm}\epsilon_{,a}+\phi e^w\delta_{,a}=0
\end{equation}
which are to be regarded as equations to find $\phi$ and
$\epsilon$ with $w$ and $\delta$ given. The general solutions are
$\phi=De^{-w}$ and $\epsilon=C-D\delta$ for $C,D\in\mathbf{R}$ and
so if $g_{ab}=De^{-w}h_{ab}+(C-D\delta)t_{ab}$, then
$g_{ab;c}=0$. Further, it is clear that $C$ and $D$ may be chosen
so that $g$ is non-degenerate at $m$ and hence on a (possibly
reduced) open neighbourhood $W$ of $m$. It is also clear that the
original null tetrad $l,n,x,y$ (with respect to $h$) is, after a
possible smooth rescaling of these tetrad members, a null tetrad
with respect to $g$ on $W$ and so $g$ is a Lorentz metric on $W$
compatible with $\Gamma$. This construction of $g$ and (10b) show
that any Lorentz metric compatible with $\Gamma$ is of this
general form on some open neighbourhood of $m$. \\
\\
{\em (iii)} The assumption that (2), (3.1) and (3.2) hold on
$\mathrm{int}C$ means that (7.1) and (7.2) also hold on
$\mathrm{int}C$ and hence that for $m\in\mathrm{int}C$, there is
an open neighbourhood $U$ of $m$ and a smooth nowhere zero vector
field $k$ on $U$ such that
\begin{eqnarray}
h_{ab;c}=h_{ab}w_c+k_ak_b\lambda_c \\
h_{ab;cd}=h_{ab}X_{cd}+k_ak_bY_{cd}
\end{eqnarray}
for necessarily smooth 1-forms $w$ and $\lambda$ and tensors $X$
and $Y$. If (23) is covariantly differentiated and equated with
(24) and the resulting equation contracted at each $m\in U$ with
$t^at^b$, where $t\in T_mM$ is such that $h_{ab}t^at^b\neq0,\
h_{ab}k^at^b=0$, one finds that on $U$
\begin{equation}
  X_{ab}=w_aw_b+w_{a;b}
\end{equation}
When this information is replaced in the equation from which it
came, one finds, after cancellation, that
\begin{equation}
  k_a(k_bY_{cd}-k_b\lambda_{c;d}-k_bw_c\lambda_d-k_{b;d}\lambda_c)=k_{a;d}k_b\lambda_c
\end{equation}
The value of $\lambda_a$ in (23) depends, of course on the choice
of $k$ and $U$ there being the freedom to replace $k$ by $\mu k$
for some nowhere zero smooth function
$\mu:U\rightarrow\mathbf{R}$. However, the condition that
$\lambda(m)=0$ is independent of the choice of $k$. So let $C_1$
(respectively $C_2$) be the subset of points of $\mathrm{int}C$
at which this condition holds (respectively, does not hold). Then
$C_2$ is open in $\mathrm{int}C$ and hence in $M$ and one has a
disjoint decomposition of $\mathrm{int}C$ in the form
\begin{equation}
\mathrm{int}C=C_1\cup C_2=\mathrm{int}C_1\cup C_2\cup
Z_1\end{equation} where $\mathrm{int}C_1$ is the interior of
$C_1$ in either the topology of $M$ or the subspace topology of
$\mathrm{int}C$ (these being equal). The subset $Z_1$ is defined
by the disjointness of the decomposition (27) and hence, since
$Z_1\subseteq C_1$, $Z_1$ has empty interior in either $M$ or
$\mathrm{int}C$. Thus $\mathrm{int}C_1\cup C_2$ is open in $M$
and open and dense in $\mathrm{int}C$. It follows from (26) that
$k$ is {\em recurrent} on $C_2$, $k_{a;b}=k_ap_b$, where $p$ is
now a smooth 1-form on $C_2$.

On $\mathrm{int}C_1$, $\lambda=0$ and so (23) gives
$h_{ab;c}=h_{ab}w_c$. A similar argument to one earlier then shows
that each $m\in\mathrm{int}C_1$ admits an open neighbourhood $W$
on which $w_a=w_{,a}$ and on which $g_{ab}=e^{-w}h_{ab}$ is then a
Lorentz metric compatible with $\Gamma$.

On $C_2$, the recurrence of $k$ together with the properties of
class $C$ and the Ricci identity give
\begin{equation}
k_{a;[bc]}=0\hspace{2cm}(\Rightarrow k_ap_{[b;c]}=0\ \ \Rightarrow
\ p_{[a;b]}=0)
\end{equation}
and so $p_a$ is locally a gradient. Thus each $m\in C_2$ admits a
neighbourhood $W$ on which $p_a=p_{,a}$ for some smooth function
$p:W\rightarrow\mathbf{R}$. Thus on $W$ the 1-form
$k'_a=e^{-p}k_a$ is covariantly constant, $k'_{a;b}=0$. Now
return to (23), which can be assumed to hold on $W$ with $k_a$
replaced by $k'_a$ (and the prime on $k'$ dropped), construct
$h_{ab;cd}$ and skew symmetrise over the indices $c$ and $d$
(using $h_{ab;[cd]}=0$ from (2)) to get
\begin{equation}
h_{ab}w_{[c;d]}+k_ak_b(\lambda_{[c;d]}-\lambda_{[c}w_{d]})=0
\end{equation}
Now one proceeds in a similar manner to that in part {\em (ii)}
to get the relations
\begin{equation}
w_{[a;b]}=0\hspace{2cm}\lambda_{[a;b]}-\lambda_{[a}w_{b]}=0
\end{equation}
From this one has on $W$ (by reducing $W$ if necessary) a
function $w:W\rightarrow\mathbf{R}$ such that $w_a=w_{,a}$ and a
relation like (20) with $\gamma_a$ replaced by $\lambda_a$ and
hence a function $\delta:W\rightarrow\mathbf{R}$ such that
$\lambda_a=e^w\delta_{,a}$. The potential metrics on $W$ are thus
of the form
\begin{equation}
g_{ab}=\phi h_{ab}+\epsilon k_ak_b
\end{equation}
for necessarily smooth functions
$\phi,\epsilon:W\rightarrow\mathbf{R}$ and with $\phi$ positive
which, from $g_{ab;c}=0$, are easily seen to satisfy the same
conditions as $\phi$ and $\epsilon$ do in (22). Thus the general
solution is $\phi=De^{-w}$ and $\epsilon=C-D\delta$ for
$C,D\in\mathbf{R}$ chosen to preserve Lorentz signature on a
(possibly reduced) open subset of $W$ (and the signature is
automatically Lorentz at those points of $W$ where $k$ is null
with respect to $h$). Thus $g$ is a Lorentz metric on $W$
compatible with $\Gamma$.\\
\\
{\em (iv)} For this part one assumes that (2) together with
(3.1), (3.2) and (3.3)hold. This is equivalent to assuming (2)
together with (7.1), (7.2) and (7.3). Thus if $m\in\mathrm{int}D$
there are smooth covector fields $r$ and $s$ defined on some open
neighbourhood $U$ of $m$ such that $r(m')$ and $s(m')$ are
independent members of $T^*_{m'}M$ at each $m'\in U$ (section 2)
and such that, on $U$,
\begin{eqnarray}
h_{ab;c}=h_{ab}w_c+r_ar_b\beta_c+s_as_b\gamma_c+2r_{(a}s_{b)}\delta_c\\
h_{ab;cd}=h_{ab}X_{cd}+r_ar_bY_{cd}+s_as_bZ_{cd}+2r_{(a}s_{b)}W_{cd}\\
h_{ab;cde}=h_{ab}X_{cde}+r_ar_bY_{cde}+s_as_bZ_{cde}+2r_{(a}s_{b)}W_{cde}
\end{eqnarray}
where $w_a,\beta_a,...W_{abc}$ are smooth tensors of the
appropriate type on $U$. Now, by shrinking $U$ if necessary, one
may assume that, in addition to the smooth covector fields $r$ and
$s$ on $U$, there are smooth covector fields $e$ and $f$ on $U$
such that $r(m'), s(m'), e(m')$ and $f(m')$ are independent
members of $T^*_{m'}M$ at each $m'\in U$. Then one defines a {\em
positive definite} metric $\gamma$ on $U$ by
$\gamma_{ab}=r_ar_b+s_as_b+e_ae_b+f_af_b$ (and where
$\gamma^{ab}$ temporarily denotes the inverse of $\gamma_{ab}$
and {\em not} indices raised with metric $h$). Then define smooth
vector fields $R, S, E$ and $F$ on $U$ by
$R^a=\gamma^{ab}r_b,...,F^a=\gamma^{ab}f_b$ and smooth covector
fields $p,q, p'$ and $q'$ on $U$ by $p_a=R^br_{b;a}$,
$q_a=S^br_{b;a}$, $p'_a=R^bs_{b;a}$ and $q'_a=S^bs_{b;a}$.
Finally define two type $(0,2)$ tensor fields $u$ and $u'$ on $U$
by
\begin{eqnarray}
r_{a;b}=r_ap_b+s_aq_b+u_{ab}\\
s_{a;b}=r_ap'_b+s_aq'_b+u'_{ab}
\end{eqnarray}
Thus $R, S, E$ and $F$ form a $\gamma$-orthogonal tetrad at each
point of $U$ and a contraction of each of (35) and (36) with $R^a$
and with $S^a$ gives
\begin{equation}
R^au_{ab}=R^au'_{ab}=S^au_{ab}=S^au'_{ab}=0
\end{equation}
Now take the covariant derivative of (32) using (35) and (36) and
equate to (33) [It is noted at this point that the above
construction of the vector fields $R$, $S$, $E$ and $F$ may be
achieved whilst, at the same time, ensuring that $E$ is nowhere
null on $U$ {\it with respect to $h$} (that is, $h_{ab}E^aE^b$ is
nowhere zero on $U$)]. A contraction of the resulting equation
with $E^aE^b$ gives
\begin{equation}
X_{ab}=w_{a;b}+w_aw_b
\end{equation}
whilst successive contractions of this same equation with
$R^aR^b$, $S^aS^b$ and $R^aS^b$, using (37) and taking into
account (38) give
\begin{eqnarray}
\nonumber
Y_{ab}=w_{a}\beta_b+2\beta_ap_b+2\delta_ap'_b+\beta_{a;b}\\
Z_{ab}=w_{a}\gamma_b+2\gamma_aq'_b+2\delta_aq_b+\gamma_{a;b}\\
\nonumber
W_{ab}=w_{a}\delta_b+\beta_aq_b+\gamma_ap'_b+\delta_aq'_{b}+\delta_ap_b+\delta_{a;b}
\end{eqnarray}
A back substitution of (39) into the equation from which they
arose and using (35) and (36) then gives after a long but
straightforward calculation
\begin{equation}
r_{(a}u_{b)d}\beta_c+s_{(a}u_{b)d}\delta_c+r_{(a}u'_{b)d}\delta_c+s_{(a}u'_{b)d}\gamma_c=0
\end{equation}
Finally, contractions of (40) with $R^a$ and $S^a$ using (37)
give, respectively, at $m$
\begin{equation}
u_{bd}\beta_c+u'_{bd}\delta_c=0\hspace{2cm}u'_{bd}\gamma_c+u_{bd}\delta_c=0
\end{equation}

Now given $m\in U$, as above, the covector fields $r$ and $s$ are
determined up to changes $r\rightarrow r'$ and $s\rightarrow s'$
where
\begin{equation}
r'=\rho r+\sigma s\hspace{2cm} s'=\rho'r+\sigma's
\end{equation}
where $\rho,\sigma,\rho'$ and $\sigma'$ are smooth functions
$:U\rightarrow\mathbf{R}$ (with $U$ possibly reduced) such that
$r'(m')$ and $s'(m')$ are independent at each $m'\in U$. The
covector field $w$ in (32) is clearly independent of the choice
of $r$ and $s$ but the covector fields $\beta,\gamma$ and
$\delta$ in (32) are not. However, the condition that
$\beta(m)=\gamma(m)=\delta(m)=0$ is independent of the choice of
$r$ and $s$. Let $D_1$ (respectively, $D_2$) be the subset if
$\mathrm{int}D$ of those points where this condition is satisfied
(respectively, not satisfied). Then $\mathrm{int}D_1$ and $D_2$
are open in $\mathrm{int}D$ and hence in $M$. Thus
$\mathrm{int}D$ admits the disjoint decomposition
$\mathrm{int}D=\mathrm{int}D_1\cup D_2\cup Z_2$ where $Z_2$ has
empty interior in $\mathrm{int}D$ and in $M$. Next, let $m\in
D_2$. The tensors $u$ and $u'$ at $m$ depend on the choice of $r$
and $s$ from (35), (36) and (42). However, it is straightforward
to check that the condition that $u(m)=u'(m)=0$ is independent of
the choice of $r$ and $s$. So let $D_3$ (respectively, $D_4$) be
the subset of $D_2$ of those points where this condition is
satisfied (respectively, not satisfied). Then $\mathrm{int}D_3$
and $D_4$ are open subsets of $D_2$ and hence of $M$ and one has
a disjoint decomposition of $D_2$ given by
$D_2=\mathrm{int}D_3\cup D_4\cup Z_3$ where $Z_3$ has empty
interior in $M$. Thus one has a disjoint decomposition of
$\mathrm{int}D$ given by
\begin{equation}
\mathrm{int}D=\mathrm{int}D_1\cup \mathrm{int}D_3\cup D_4\cup Z_4
\end{equation}
where $Z_4=Z_2\cup Z_3$. Now $Z_3\subseteq D_2$ and $Z_2\cap
D_2=\emptyset$ and so if $V(\neq\emptyset)$ is open in $M$ with
$V\subseteq Z_4$ then, since $Z_2$ and $Z_3$ have empty interiors
in $M$, $V$ cannot be contained in either $Z_2$ or $Z_3$ and hence
$V\cap Z_2\neq\emptyset\neq V\cap Z_3$. But since $D_2$ is open,
$V\cap D_2$ is open and a subset of $Z_3$ and is hence empty. Now
$V\cap Z_3\subseteq V\cap D_2$ and so one achieves the
contradiction that $V\cap Z_3=\emptyset$. It follows that
$V=\emptyset$ and so $Z_4$ has empty interior in $M$.

Let $m\in\mathrm{int}D_1$ so that, from (32),
$h_{ab;c}=h_{ab}w_c$ on some open neighbourhood $U$ of $m$. Then,
by reducing $U$ if necessary, one has, as before (see (12)), a
smooth function $w:U\rightarrow\mathbf{R}$ such that $w_a=w_{,a}$
and then $e^{-w}h_{ab}$ is a Lorentz metric on $U$ compatible with
$\Gamma$.

Let $m\in\mathrm{int}D_3$ so that, whichever covector fields $r$
and $s$ are chosen, (35) and (36) hold on some open neighbourhood
$U$ of $m$ with $u$ and $u'$ zero on $U$. Then it follows from
these equations and the definition of the curvature class $D$ that
\begin{eqnarray}
\nonumber r_aR^a_{\ bcd}=0,\ \ \ r_aR^a_{\ bcd;f_1...f_n}=0\ \ \
\ (n=1,2,...)\\
\\
\nonumber s_aR^a_{\ bcd}=0,\ \ \ s_aR^a_{\ bcd;f_1...f_n}=0\ \ \
\ (n=1,2,...)
\end{eqnarray}
Thus the infinitesimal holonomy group of $\Gamma$ is
1-dimensional at each $m'\in U$ and, given $U$ is chosen
connected and simply-connected (as it always can be), is
isomorphic (as a Lie group) to the holonomy group of $\Gamma$
(restricted to $U$) and with holonomy algebra represented by the
matrices in the range of the map $f$ described in section 2.
Since (2) holds this holonomy group is a subgroup of the
orthogonal group of $h$ and it follows that a metric on $U$
compatible with $\Gamma$ is now assured \cite{Schmidt} (see, also
\cite{K&N}) . To actually construct such a metric one notes that
the members of $T_mM$ associated with $r(m)$ and $s(m)$ under $h$
annihilate each matrix in the above representation of the
holonomy algebra. Hence, by exponentiation, they give rise to
holonomy invariant (i.e. covariantly constant) vector fields on
$U$ which span the same distribution on $U$ as do the vector
fields associated with $r$ and $s$ under $h$ (see, e.g.
\cite{K&N,HallBook}). Hence one may now suppose, using the freedom
(42), that $r$ and $s$ above are covariantly constant. In this
case, covariantly differentiating (32) and using the Ricci
identity for $h$ and (2), one finds
\begin{equation}
  w_{[a;b]}=0,\ \ \ \beta_{[a;b]}=\beta_{[a}w_{b]},\ \ \ \gamma_{[a;b]}=\gamma_{[a}w_{b]},
  \ \ \ \delta_{[a;b]}=\delta_{[a}w_{b]}
\end{equation}
and so since $U$ is simply connected, $w_a=w_{,a}$ with
$w:U\rightarrow\mathbf{R}$. Then (45) shows that $e^{-w}\beta$,
$e^{-w}\gamma$ and $e^{-w}\delta$ are closed 1-forms on $U$ and
so there exist $\mu,\nu,\lambda:U\rightarrow\mathbf{R}$ such that
$\beta_a=e^w\mu_{,a}$, $\gamma_a=e^w\nu_{,a}$ and
$\delta_a=e^w\lambda_{,a}$. Now construct the tensor $g$ on $U$ by
\begin{equation}
g_{ab}=e^{-w}h_{ab}+(c_1-\mu)r_ar_b+(c_2-\nu)s_as_b+2(c_3-\lambda)r_{(a}s_{b)}
\end{equation}
where $c_1,c_2,c_3\in\mathbf{R}$. It is easily checked that $g$
is covariantly constant and, by appropriate choice of $c_1,c_2$
and $c_3$ (e.g. $c_1(m)=\mu(m)$, $c_2(m)=\nu(m)$ and
$c_3(m)=\lambda(m)$), $g$ can be made to have Lorentz signature in
some neighbourhood of $m$.

Now let $m\in D_4$ so that at least one of $\beta(m)$,
$\gamma(m)$ and $\delta(m)$ is not zero. Then, by using the
freedom permitted by (42), one may arrange that $\beta$, $\gamma$
and $\delta$ are each non-zero at $m$ and hence in some open
neighbourhood $U\subseteq D_4$ of $m$. It then follows from (41)
and the definition of the region $D_4$ that $u$ and $u'$ are
nowhere zero and proportional on $U$ so that, on $U$, $u'=\kappa
u$ with $\kappa:U\rightarrow\mathbf{R}$ nowhere zero. Now one can
again use the freedom in (42) by replacing $r$ by $r'=\kappa r-s$
(with $s$ unchanged) to achieve (35) and (36) with $u$ zero and
$u'$ nowhere zero on $U$. It then follows from (41) that $\gamma$
and $\delta$ vanish on $U$ (and hence that $\beta$ is nowhere
zero on $U$). Then (32) is
\begin{equation}
h_{ab;c}=h_{ab}w_c+r_ar_b\beta_c
\end{equation}
where, for convenience, the prime on $r$ is omitted. From (47) and
(35) (with $u=0$) one then finds
\begin{equation}
h_{ab;cd}=h_{ab}(w_{c;d}+w_cw_d)+r_ar_b(\beta_{c;d}+w_c\beta_d+2\beta_cp_d)+2r_{(a}s_{b)}\beta_cq_d
\end{equation}
At this point the extra condition (3.3) (or, equivalently (7.3)
or (34)) is introduced. Thus one takes one more covariant
derivative of (48) and equates it to the right hand side of (34)
to obtain on equation on $U$ of the form
\begin{equation}
h_{ab}A_{cde}+r_ar_bB_{cde}+s_as_bC_{cde}+2r_{(a}s_{b)}D_{bde}+2r_{(a}u'_{b)e}\beta_cq_d=0
\end{equation}
where the tensors $A,B, C$ and $D$ can be calculated but whose
exact form is not needed. A contraction with $E^aE^b$ (cf. the
proof of (38)) gives $A=0$ on $U$ whilst contractions with
$R^aR^b$, $S^aS^b$ and $R^aS^b$ reveal that $B=C=D=0$ on $U$.
Thus, on $U$, (49) becomes
\begin{equation}
r_{a}u'_{be}\beta_cq_d+r_{b}u'_{ae}\beta_cq_d=0
\end{equation}
A contraction of (50) with $R^a$ and the use of (37) then reveals
that $q_a=0$ on $U$ (since $u'$ and $\beta$ are nowhere zero on
$U$). Thus from (35), $r_{a;b}=r_ap_b$ and then the Ricci
identity on $r$ reveals that $p_{[a;b]}=0$ and so, by shrinking
$U$ if necessary, one has $p_a=p_{,a}$ for some function
$p:U\rightarrow\mathbf{R}$. Then one can replace $r$ in the above
by $e^{-p}r$ where $(e^{-p}r_a)_{;b}=0$. This latter (covariantly
constant) vector field will still be labelled $r$. Then (47) and
(2) show that $w_{[a;b]}=0$ and (similar to case {\em (iii)})
$\beta_{[a;b]}-\beta_{[a}w_{b]}=0$. Thus, again by reducing $U$
if necessary one has functions $w,\theta:U\rightarrow\mathbf{R}$
such that $w_a=w_{,a}$ and $\beta_a=e^w\theta_{,a}$. Then tensor
$g$ defined on $U$ by
\begin{equation}
g_{ab}=e^{-w}h_{ab}-(\theta+C)r_ar_b
\end{equation}
is now easily checked to be covariantly constant, where the
arbitrary constant $C$ can be adjusted to ensure that $g$ has
Lorentz signature in some open neighbourhood of $m$.

In summary, the manifold $M$ has been disjointly decomposed as
$M=M'\cup Z'$ where $M'$ is the open subset of $M$ given by
\begin{equation}
M'=A\cup\mathrm{int}B_1\cup\mathrm{int}B_2\cup\mathrm{int}C_1\cup
C_2\cup\mathrm{int}D_1\cup\mathrm{int}D_3\cup D_4\cup\mathrm{int}O
\end{equation}
each point of which (under the appropriate conditions in theorem
2) admits an open neighbourhood and a metric on that neighbourhood
compatible with $\Gamma$ (this being clearly true with no further
conditions on the region $\mathrm{int}O$) and where $Z'=Z\cup
Z_1\cup Z_4$. Now each of $Z$, $Z_1$ and $Z_4$ has empty interior
in $M$ and $Z$ is closed in $M$. Also $Z_1\subseteq\mathrm{int}C$
and $Z_4\subseteq\mathrm{int}D$ and so (since, from (6), $Z$ is
disjoint from $\mathrm{int}C$ and $\mathrm{int}D$) $Z$, $Z_1$ and
$Z_4$ are mutually disjoint. It now follows that $Z'$ is a closed
subset of $M$ with empty interior and hence that $M'$ is open and
dense in $M$. To see this one first notes that $Z'$ is closed by
definition and that from the preliminary decompositions (6) and
(27), $Z\cup Z_1$ is closed. Now let $\emptyset\neq V\subseteq
Z'$ with $V$ open. Then $V'\equiv V\cap(M\backslash Z)$ is open
and $V'\cap Z=\emptyset$. Then $V''\equiv
V'\cap(M\backslash(Z\cup Z_1))$ is open and $V''\subseteq Z_4$.
Thus $V''=\emptyset$ and hence $V'\subseteq Z\cup Z_1$. Thus
$V'\subseteq Z_1$ and so $V'=\emptyset$. This means that
$V\subseteq Z$ which is a contradiction. Thus
$\mathrm{int}Z'=\emptyset$ and $M'$ is open and dense in $M$.
This completes the proof of theorem 2. $\bullet$

The next two theorems are consequences of theorem 2.

\begin{theorem}
Under the conditions of theorem 2, if $\Gamma$ is Ricci flat then
$D=\emptyset$ and equations (2), (3.1) and (3.2) on $M$ are
sufficient to ensure a local Lorentz metric compatible with
$\Gamma$ on the open subset $M'$ of $M$.
\end{theorem}
{\bf Proof}\\ If the curvature class at $m\in M$ is $D$ then the
curvature tensor takes the form $R_{abcd}(=h_{ae}R^e_{\
bcd})=\lambda F_{ab}F_{cd}$ for some simple bivector $F$ at $m$
and $0\neq\lambda\in\mathbf{R}$ (section 2). The Ricci flat
condition $R^c_{\ bcd}=0$ then gives $F^c_{\ b}F_{cd}=0$. Since
$F$ is simple one may write $F=p\wedge q$ for $p,q\in T^*_mM$ and
with $p$ and $q$ ($h$-)orthogonal. The Ricci flat condition then
shows that $p$ and $q$ are each ($h$-)null, contradicting the
Lorentz signature of $h$. Thus $D=\emptyset$ and the result now
follows from theorem 2. $\bullet$

\begin{theorem}\hspace{12cm}
\begin{enumerate}
\item If $m\in M'$, with $M'$ given by (52) and if $h$ satisfies
(2) and those equations in (3) appropriate to the region of $m'$
in which $m'$ is located, and as given in theorem 2, in that
region, then $h$ satisfies (3.$k$) for all $k$ at each point of
that region.
\item If the infinitesimal holonomy algebra associated with
$\Gamma$ has constant dimension on $M$ and $M$ is simply
connected and if the conditions appropriate for each of the
regions as specified in theorem 2 hold in these regions, $M$
admits a global Lorentz metric which is compatible with $\Gamma$.
\end{enumerate}
\end{theorem}
{\bf Proof}\\ {\em (i)} The proof of this follows rather quickly
by using the various relationships obtained in the proof of
theorem 2 between $h$ and the constructed local metric $g$. Thus
either $h$ and $g$ were conformally related or linked by various
recurrent tensors $t$ or recurrent 1-forms $k$. The result now
follows from (1).\\
\\ {\em (ii)}
The constancy of the dimension of the infinitesimal holonomy
algebra means that the infinitesimal holonomy groups at each
$m\in M$ are isomorphic to each other and to the restricted
holonomy group of $M$ and hence (since $M$ is simply connected)
to the holonomy group of $M$ \cite{K&N}. Then part (i) shows that
$h$ satisfies (3.$k$), for all $k$, on $M'$ and hence on $M$.
Thus the infinitesimal holonomy algebra at each $m\in M$ is a
subalgebra of the orthogonal algebra of $h$ (since each
contribution to the infinitesimal holonomy algebra satisfies
(4)). It follows that the holonomy group of $M$ is isomorphic to a
subgroup of the orthogonal group of $h$ (that is, its members
preserve the quadratic form $h$ at each $m\in M$ \cite{Schmidt} -
see, also, \cite{K&N} and section 1). The result in (ii) now
follows. (In fact the original conditions (2) and (3.1)-(3.$k$)
imposed may be regarded as forcing a certain subspace of the
infinitesimal holonomy algebra to lie within the orthogonal
algebra of $h$.) $\bullet$

\section{Remarks and Examples}
In this paper the tensor $h$ was assumed to be of Lorentz
signature. However it is clear that if $h$ is assumed
positive-definite, the curvature types $A$, $B$, $C$, $D$ and $O$
still make sense after an obvious change for type $B$ where the
range of $f$ would be 2-dimensional and spanned, in an obvious
notation, by $x^ay_b-y^ax_b$ and $z^aw_b-w^az_b$ where $x,y,z,w$
is an orthonormal tetrad at the point in question. Also, even
when $h$ is taken as having Lorentz signature, it has been
pointed out that metrics of both Lorentz {\em and}
positive-definite signature have sometimes been constructed and
which are compatible with $\Gamma$ (cf. \cite{HallBook}).

The results of theorem 2 give sufficient conditions in terms of
equations (2) and (3.1)-(3.$k$) for each curvature class for a
connection to be metric. The work of Edgar shows that for a
fairly general class of cases the condition (2) is sufficient,
but the class to which this applies is somewhat obscure
\cite{Edgar}. Here some examples will be given which show that
condition (2) is not a sufficient condition for the existence of
a local metric for any of the curvature classes $A$, $B$, $C$ or
$D$.\\

{\em Example 1}. This example is taken from
\cite{Hall&Haddow1995}. Let $(\mathrm{R}^4,\eta)$ denote
Minkowski space and let $U$ be the open submanifold $t>0$ of
$\mathrm{R}^4$. Define $\phi:U\rightarrow\mathrm{R}$ by
$\phi(x,y,z,t)=\log t$ so that $\phi_{a;b}=-\phi_a\phi_b$ where
$\phi_a=\phi_{,a}$. Define a metric $g$ on $U$ by
$g=e^{-2\phi}\eta$ and let $\Gamma$ and $R$ be the associated
Levi-Civita connection and curvature tensor, respectively. Then
in the coordinates on $U$ inherited from $\mathrm{R}^4$
\begin{eqnarray}
\Gamma^a_{\ bc}=-\left(\delta^a_{\ b}\phi_c+\delta^a_{\
c}\phi_b-\phi^ag_{bc}\right) \\
R^a_{\ bcd}=\phi^e\phi_e\left(\delta^a_{\ d}g_{bc}-\delta^a_{\
c}g_{bd}\right)
\end{eqnarray}
where $\phi^a=g^{ab}\phi_b$. Then define a 1-form $\psi$ on $U$
by $\psi_a=(1-e^{\phi})^{-1}\phi_a$ and a new symmetric connection
$\Gamma'$ on $U$ by the components (cf. \cite{Schouten})
\begin{equation}
\Gamma'^a_{\ bc}=\Gamma^a_{\ bc}+\delta^a_{\ b}\psi_c+\delta^a_{\
c}\psi_b
\end{equation}
This has the curvature tensor $R'$ with components
\begin{equation}
R'^a_{\
bcd}=\phi^e\phi_e\left(1-(1-e^{\phi})^{-1}\right)\left(\delta^a_{\
d}g_{bc}-\delta^a_{\ c}g_{bd}\right)
\end{equation}
It is easy to check that $R'$ and $g$ satisfy (2) and that the
map $f$ associated with $R'$ (section 2) has rank equal to 6
everywhere on $U$ and so the curvature class is everywhere $A$.
However, using $R'$ and $\Gamma'$ one can check after a short
calculation that (3.1) fails. In fact $\Gamma'$ is not a local
metric connection because if it was compatible with a metric $g'$
on some open subset $V$ of $U$ then from (10a) one has without
loss of generality $g'=e^{\alpha}g$ for some smooth function
$\alpha:V\rightarrow\mathrm{R}$. But $g$ has zero covariant
derivative with respect to $\Gamma$ and so the condition that
$g'$ has zero covariant derivative with respect to $\Gamma'$
gives, using (55)
\begin{equation}
\alpha_{,c}g_{ab}-2g_{ab}\psi_c-g_{ac}\psi_b-g_{bc}\psi_a=0
\end{equation}
On contracting (57) with $g^{ab}$ one finds $2\alpha_{,c}=5\psi_c$
and a back substitution gives
$g_{ab}\psi_c=2g_{ac}\psi_b+2g_{bc}\psi_a$. A final contraction
at any $p\in V$ with $X^c$ for any $X\in T_pV$ and a
consideration of rank reveals that $X^a\psi_a=0$ for all such $X$
and hence the contradiction that $\psi\equiv0$ on $V$. Thus
although $\Gamma'$ yields a curvature tensor of class $A$
satisfying (2), (3.1) fails and $\Gamma'$ is {\em not} locally
metric.\\

{\em Example 2}. This example is quite general and can be applied
to each of the curvature classes $A$, $B$, $C$ and $D$. Let
$(M,g)$ be a space-time which admits a nowhere zero {\em
recurrent null} vector field $l$ with the property that no
rescaling of $l$ is covariantly constant over some non-empty open
subset of $M$. This means that at some $p\in M$, $R^a_{\
bcd}l^d\neq0$, for otherwise, one has $l_{a;b}=l_ap_b$ for some
covector field $p$ on $M$ which from the Ricci identity satisfies
$p_{[a;b]}=0$ on $M$. Thus for any $p\in M$ there is an open
neighbourhood $U$ of $p$ and a function
$\phi:U\rightarrow\mathrm{R}$ such that $p_a=\phi_{,a}$ and then
$e^{-\phi}l$ is a covariantly constant rescaling of $l$ over $U$.
Thus there exists $p\in M$ and, by continuity, some open
neighbourhood $V$ of $p$ on which $R^a_{\ bcd}l^d$ is nowhere
zero. The recurrence condition and the Ricci identity on $l$ in
fact show that $R^a_{\ bcd}l^d=F^a_{\ b}l_c$ for some tensor $F$
on $V$ which is hence nowhere zero on $V$. Restricting attention
to $V$ with the original metric $g$ and associated Levi-Civita
connection $\Gamma$ it is noticed that $l_{[a}l_{b;c]}=0$ and
hence that $l$ is hypersurface orthogonal on $V$. Thus, shrinking
$V$, if necessary, one may assume $l$ is scaled to a gradient
vector field $l'$ on $V$ such that $l'_{a;b}=\psi l'_al'_b$ for
some function $\psi:V\rightarrow\mathrm{R}$. Now define a new
symmetric connection $\Gamma'$ on $V$ by the components
$\Gamma'^a_{\ bc}=\Gamma^a_{\ bc}+l'^al'_bl'_c$. It is easily
calculated that the curvature tensor $R'$ on $V$ arising from
$\Gamma'$ is identical to the original one (this can be quickly
checked by first noting that if $P^a_{\ bc}=\Gamma'^a_{\
bc}-\Gamma^a_{\ bc}=l'^al'_bl'_c$ then $P^a_{\ bd;c}-P^a_{\
bc;d}=0$ and $P^a_{\ bc}P^c_{\ de}=0$ and then using a convenient
formula in \cite{Eisenhart,Rosen}). Thus in any chart on $V$,
\begin{equation}
R'^a_{\ bcd}=R^a_{\ bcd}
\end{equation}
Using a vertical stroke to denote a $\Gamma'$ covariant
derivative one thus has on $V$, using another convenient result
in \cite{Eisenhart,Rosen},
\begin{eqnarray}
\nonumber R'^a_{\ bcd|e} &=R^a_{\ bcd|e}\\
&=R^a_{\ bcd;e}+R^f_{\ bcd}P^a_{\ fe}-R^a_{\ fcd}P^f_{\
be}-R^a_{\ bfd}P^f_{\ ce}-R^a_{\ bcf}P^f_{\ de}\\
\nonumber &=R^a_{\ bcd;e}-2l'^al'_bl'_eF_{cd}
\end{eqnarray}
It now easily follows that for $R'^a_{\ bcd}$, $R'^a_{\ bcd|e}$
and $g$, (2) holds and (3.1) fails on $V$.

Several cases of these results can now be considered which cover
each of the curvature classes $A$, $B$, $C$ and $D$ and for none
of which $\Gamma'$ is locally metric.

Consider first the situation where the original $g$ is a vacuum
metric of the Goldberg-Kerr type
\cite{Goldberg&Kerr,Kerr&Goldberg}. Then the associated space-time
may be taken as $V$ and admitting a recurrent null vector field
$l$ as described above. The Petrov type is III and so the
curvature tensor (being equal to the Weyl tensor because of the
vacuum condition) has an associated map $f$ of rank 4 (see, e.g.
\cite{HallBook}). The curvature class is thus $A$. If $\Gamma'$
defined as above was compatible with a metric $g'$ on some open
subset $W$ of $V$ then $\Gamma$ and $\Gamma'$ are each metric
connections on $W$ with the same curvature tensor. It then
follows from \cite{Hall1983,HallBook} that, on $W$, $g'=cg$
$(0\neq c\in\mathrm{R})$ and hence the contradiction that
$\Gamma'=\Gamma$ on $W$.

Now consider the situation when $V$ is a space-time of curvature
class $B$ and hence of holonomy type $R_7$
\cite{HallBook,Hall&Lonie2000} and thus the metric $g$ is of the
Bertolli-Robinson type (see, e.g. \cite{KramerEtAl}). Further, if
$V$ is chosen to be simply connected (as it always can) then two
independent recurrent null vector fields $l$ and $n$ are admitted
by $V$ with each having all of the properties that $l$ had in the
previous paragraph and satisfying $l^an_a=1$, $l_{a;b}=l_ap_b$
and $n_{a;b}=-n_ap_b$ \cite{HallBook,Hall&Kay1988,Hall&Lonie2000}.
Thus the tensor $h$ on $V$ defined by $h_{ab}=2l_{(a}n_{b)}$ is
covariantly constant on $V$, $h_{ab;c}=0$. Now define a symmetric
connection $\Gamma'$ by the components $\Gamma'^a_{\
bc}=\Gamma^a_{\ bc}+l'^al'_bl'_c$ and suppose that $\Gamma'$ is
compatible with a metric $g'$ on some open subset $W$ of $V$.
Then $\Gamma$ and $\Gamma'$ give the same curvature tensor as
before and it follows from
\cite{HallBook,Hall&McIntosh1983,Hall&Kay1988} (see (10b)) that
$g'=\phi g+\psi h$ for functions
$\phi,\psi:W\rightarrow\mathrm{R}$. Imposing the condition
$g_{ab;c}=0=g'_{ab|c}$, recalling that $h_{ab;c}=0$ and taking
$P^a_{\ bc}=l'^al'_bl'_c$ as before gives
\begin{eqnarray}
\nonumber 0=g'_{ab|c}&=\phi_cg_{ab}+\phi g_{ab|c}
+\psi_{,c}h_{ab}+\psi h_{ab|c} \\
\nonumber &=\phi_{,c}g_{ab}+\phi\left(g_{ab;c}-g_{ae}P^e_{\
bc}-g_{eb}P^e_{\ ac}\right)\\
\nonumber &\hspace{5mm}
+\psi_{,c}h_{ab}+\psi\left(h_{ab;c}-h_{ae}P^e_{\ bc}-h_{eb}P^e_{\
ac}
\right)\\
\Rightarrow &
\left(\phi+\psi\right)_{,c}g_{ab}=2(\phi+\psi)l'_al'_bl'_c
\end{eqnarray}
A rank argument shows that $\phi=-\psi$ and hence that
$g'=\phi(g-h)$. But then $g'_{ab}l^b=g'_{ab}n^b=0$ and so $g'$ is
not non-degenerate. This contradiction completes the proof.

The final two cases may be dealt with briefly. First, let $(M,g)$
be a space-time of holonomy type $R_6$
(\cite{HallBook,Hall&Kay1988,Hall&Lonie2000} - and for existence
and examples see \cite{Debever&Cahen}). Then, restricting to an
open subset $U$ of $M$, if necessary, one may assume that, on
$U$, there are vector fields $y$ and $l$ with $y$ spacelike and
covariantly constant and $l$ null and recurrent. Thus $y_{a;b}=0$
and $l_{a;b}=l_ap_b$ with $p$ a nowhere zero 1-form on $U$ and one
may assume $U$ is chosen so that the map $f$ (section 2) has
everywhere rank 2. Then, on $U$, $R^a_{\ bcd}y^d=0$ (that is, the
curvature class is $C$) and $R^a_{\ bcd}l^d$ is nowhere zero. On
scaling $l$ to $l'$ as in the previous examples and constructing
the connection $\Gamma'$ as before, one again achieves (58) and
(59) so that $g$, $\Gamma'$ and $R'$ satisfy (2) but not (3.1).
Because of (58) it follows from (10c)
\cite{HallBook,Hall&McIntosh1983,Hall1983} that if $g'$ is a
metric compatible with $\Gamma'$, one must have $g'_{ab}=\phi
g_{ab}+\psi y_ay_b$ for functions
$\phi,\psi:U\rightarrow\mathbf{R}$. Now the condition
$g'_{ab|c}=0$ yields, by an argument which is very similar to
(60), the consequence that $\phi\equiv 0$ on $U$ and so that such
a tensor $g'$ is not non-degenerate. Hence $\Gamma'$ is not
locally a metric connection.

Finally for a space-time of holonomy type $R_2$
\cite{HallBook,Hall&Kay1988,Hall&Lonie2000,Debever&Cahen} one may
arrange an open subset $U$ of it to admit vector fields $l$, $n$,
$x$ and $y$ with all inner products between them vanishing on $U$
except $l^an_a=x^ax_a=y^ay_a=1$. Also, on $U$, $l$ and $n$ are
(null and) recurrent and $x$ and $y$ are covariantly constant.
The set $U$ may be chosen so that the map $f$ has rank 1
everywhere on $U$ (and hence, the curvature class is $D$). One
then proceeds as above to construct $\Gamma'$ (using $l$ or $n$)
so that $g$, $\Gamma'$ and $R'$ satisfy (2) but not (3.1) and
notes that, from (58), any metric $g'$ compatible with $\Gamma'$
satisfies \cite{HallBook,Hall&McIntosh1983,Hall1983} $g'_{ab}=\phi
g_{ab}+\alpha x_ax_b+\beta y_ay_b+2\gamma x_{(a}y_{b)}$ for
functions $\phi,\alpha,\beta,\gamma:U\rightarrow\mathbf{R}$.
Another argument, very similar to (60) then shows that
$\phi\equiv 0$ on $U$ and thus $g'$ is not non-degenerate. Hence
$\Gamma'$ is not locally a metric connection.\\

As a final remark, it is noted how the tensor $h$ and any local
metric compatible with the connection $\Gamma$ are related
geometrically. For the case relevant to theorem 2(i), that is,
class $A$, any local metric is just a conformal rescaling of $h$.
For class $B$, any local metric is a (functional) combination of
$h$ and the local covariantly constant tensor $t$, where the
latter tensor's eigenspaces (with respect to $h$) are, at each
relevant point $m$, identical to the two 2-dimensional subspaces
of $T_mM$ into which $T_mM$ is decomposed by the curvature tensor
for this class. For class $C$, any local metric is either a
rescaling of $h$ or a (functional) combination of $h$ and the
covariantly constant tensor $k_ak_b$ where $k$ is a local
covariantly constant vector field spanning the unique direction
picked out at each relevant point by a class $C$ curvature
tensor. For class $D$, any local metric $g$ is either a rescaling
of $h$ or a (functional) combination of $h$ and appropriate
products of either one or two local covariantly constant vector
fields and where these latter vector fields lie in the
2-dimensional subspace determined at each appropriate point by
this curvature class. In addition, it is noted that if $g$ and
$g'$ are two local metrics compatible with $\Gamma$ in some
neighbourhood $U$, the latter being contained in one of the
regions $A$, $\mathrm{int}B_1$, $\mathrm{int}B_2$ etc. considered
above, then $g$ and $g'$, having the same Levi-Civita connection,
are either conformally related with a constant conformal factor
or are jointly related to the appropriate covariantly constant
tensor(s) mentioned above by simple (constant coefficient)
combinations and which reflect the local holonomy group $[4,3]$.

\end{document}